\documentclass{article}
\usepackage{amsmath}
\usepackage{hyperref}
\usepackage{graphicx}
\usepackage{xcolor}
\usepackage{tikz}
\newcommand{\xlink}{  
    \tikz[x=1.2ex, y=1.2ex, baseline=0ex]{%
        \begin{scope}[x=1.3ex, y=1.3ex]
            \clip (-0.1,-0.1) 
                --++ (-0, 1.2) 
                --++ (0.6, 0) 
                --++ (0, -0.6) 
                --++ (0.6, 0) 
                --++ (0, -1);
            \path[draw, 
                line width = 0.75, 
                rounded corners=0.5, blue] 
                (0,0) rectangle (1,1);
        \end{scope}
        \path[draw, line width = 0.75, blue] (0.5, 0.5) 
            -- (1.3, 1.3);
        \path[draw, line width = 0.75, blue] (0.8, 1.3) 
            -- (1.3, 1.3) -- (1.3, 0.8);
        }
    }

\title{City size distributions are driven by each generation’s stay-or-leave decision}
\author{Robin W. Spencer\footnote{The author is a retired biochemist and data analyst.  Following degrees in physics (BA Williams College) and biochemistry (Ph.D. MIT), he worked at Syntex, Pfizer, and Imaginatik.  His current principal interests are research and software development in genetic genealogy.  Studies of monoparental DNA haplogroup migration and large genealogic databases led to these questions of city demography.  Contact spencerrw@alum.mit.edu, website \url{http://scaledinnovation.com} }}
\begin{document}
\sloppy   
\maketitle

\begin{abstract}Throughout history most young adults have chosen to live where their parents did while a smaller number moved away.  This is sufficient, by proof and simulation, to account for the well-known power law distributions of city sizes. The model needs only two parameters, $r$ = the probability that a child stays, and the maximum number of cities (which models the observed saturation at high city rank). The power law exponent follows directly as $\alpha = 1 + 1/r$, with Zipf’s Law simply the limiting case of $r \rightarrow 1$.  Observed exponents $(\alpha = 2.2 \pm 0.4, n = 158)$ are consistent with stay-or-leave data from large genealogic studies.  This model is self-initializing and could have applied from the time of the earliest stable settlements.  The driving narrative behind city-size distributions is fundamentally about family ties, familiarity, and risk-avoidance, rather than economic optimization.
\end{abstract}

\section*{Introduction}
\begin{center}
\textcolor{gray}{Insert Figure 1 here.}
\end{center}

Over a century ago, Felix Auerbach noted that the populations of German cities had a regular pattern that closely fit a power law distribution$^{2}$; his 1910 data are shown in Figure 1.  Since then his observation has been repeated hundreds of times, has spawned a huge literature in economics and mathematics, and has been seriously considered by Nobel Laureates$^{3, 4}$.  Yet explanatory theories remain incomplete and unsatisfactory as Paul Krugman noted in 1996: ``while there must be a compelling explanation of the astonishing empirical regularity in question, I have not found it."$^{3}$.

    To be considered against this impressive literature, any new proposal should address the following:
        \begin{enumerate}
            \item Human Behavior: The proposal must have a common-sense narrative for the human behavior that drives the city-size distribution, and that behavior should apply across centuries and cultures; i.e. it should not depend on modern transportation efficiency or any particular political system.
            \item Mathematical Model: It must have a generative theory and appropriate mathematics that give rise to the observed distributions.
            \item Full Range: It must provide reasonably accurate matches to datasets over their entire ranges and with a minimum of fitted parameters.  Each such parameter should not be just for fitting purposes, but have a physical interpretation appropriate to the model and human narrative.
            \item Dynamics: It must explain and match the observed dynamics of town and city growth over time.  Extrapolations to early and late times should be historically reasonable. Simulation should reproduce observed patterns over time.
            \item Initialization: It must have a simple and common-sense initialization method, i.e. the model must start at time zero when there are no cities.
            \item Limits and Exceptions: A model need not apply to all settlement and migration behavior in all eras, though its limits should be reasonable and ideally apparent in the data. Specific datasets that strongly disagree with the model should not be ignored but should have explanatory, perhaps testable, hypotheses.
            \item Clear about Alternatives: Alternative theories of city growth should be clearly identifiable as either consistent or not, and in the particulars -- for example, another theory may agree in the steady state but not at early or late times.
            \item Additional Observations: The proposal should suggest additional questions and research avenues with testable consequences.
        \end{enumerate}
    The following sections discuss how the proposed model addresses these criteria, after which follow additional questions and areas of continuing research. Note that the term ``city" is used broadly to refer to a fixed population center of any size, thus including villages and towns.
\pagebreak
    
\section*{Human Behavior}
\begin{center}
\textcolor{gray}{Insert Figure 2 here.}
\end{center}

Most of the explanatory narratives in the literature are economic or geographic in which \textit{Homo economicus} is a rational, informed, economic optimizer$^{5}$.  Here we follow instead the ideas of Kahneman and Tversky in which major decisions are made by gut feel, social forces, and with a bias against risk or loss$^{6}$.  The relevant decision is that faced by all young adults: whether to stay at or near their parents' homestead or to strike out on their own.  This is illustrated in Figure 2 in which the next generation stays with probability $r$ or leaves with probability $1 - r$.
\begin{center}
\textcolor{gray}{Insert Figure 3 here.}
\end{center}

 Supportive data are available in large public genealogy databases.  Kaplanis \textit{et al} 2018$^{7}$ analyzed 86 million genealogy profiles; Figure 3 (left) from their data shows that the trimmed mean (winsorized central 60\%) distance between a European, British, or Irish person's birth and death location was 3-10 km for those born before 1800.  Figure 3 (right) shows the stay-or-leave decision over time: until 1830-1870 birth cohort, 60-80\% of Europeans never left home, corresponding in this model to $\alpha$ = 2.3 to 2.7.  North America has been more mobile, with birth-death distances over 100 km for those born after 1740, who also had only a 25\% likelihood of staying where they were born.

\begin{center}
\textcolor{gray}{Insert Figure 4 here.}
\end{center}

    Figure 4 shows data from a similar study of internal French migration for three generations starting 1800-1804$^{8}$: the data fit a simple kinetic model in which 62\% of each generation stay at home while 38\% move 20 km or more away.   Similarly in 1885 Ravenstein found that 75\% of British and Irish resided in their birth county$^{9}$.  ``Stay local" factors of 62\%-75\% correspond to $\alpha$ = 2.3 to 2.6.

    Certainly the stay-or-leave decision would be influenced by resources and opportunity (insufficient = push, excess = pull).  For most of human history the total fertility rate TFR$^{10}$ has been only slightly greater than two \href{http://scaledinnovation.com/gg/gg.html?rr=gwatson}{\xlink} : on average two children replace two parents.  Barring external change (drought, fire, famine, war), the next generation need not move but simply occupy the farm/village/town of their forebears.  With TFR marginally above two the result is Zipf's Law$^{11}$ with $\alpha$ $\approx$ 2.

    It's also entirely possible that stay-or-leave is driven by the stochastic nature of birth and death:  in a small family adult children could stay, but in large family more may need to move away$^{12}$.  Since family sizes are Poisson distributed$^{13}$, a few large families will always occur, and simulation shows that they alone can provide the necessary seeding to break the paradox that a rigid interpretation of Zipf's exponent leads to a single massive city.

    Interestingly and simplistically, if all locations are equally attractive but people weigh losses twice as much as gains$^{6}$, then the probability to stay-or-leave should be 2/3 which gives exponent $\alpha$ = 2.5.

\section*{Mathematical Model}
\begin{center}
\textcolor{gray}{Insert Figure 5 here.}
\end{center}

 The model is taken from Bagrow 2008$^{14}$ and described \href{http://scaledinnovation.com/gg/gg.html?rr=frontierDemographics}{\xlink} in terms of a checkerboard: at each time step, with probability $r$, a new checker is placed on a square proportionate to the number of checkers already there; this is the rich-get-richer Pareto process.  But also with probability $1 - r$, a new checker is placed on any square at random, occupied or not.  This is extremely close to Simon's model, described by Krugman as ``lumps" of population making a stay-or-move decision$^{3}$, and known to generate power laws.  Bagrow adds an essential element to this highly regarded model by adapting it to a finite space -- the checkerboard, or in this context, a country or continent.  Bagrow shows that the distribution evolves as a power law at early times and for an infinite checkerboard (with exponent $\alpha = 1 + 1/r)$, but saturates in a finite space.  Once all of the squares are heavily occupied, the process becomes poisson and thus approaches a gaussian distribution, which must tail down since it is anchored -- for a literal checkerboard at rank = 64.

 Though Bagrow \textit{et al} establish early and late functional forms (power law and gaussian, respectively), they do not offer an intermediate formulation, and it may be that there is not a single simple form. If we adopt the model rigidly (there must be a finite and integer number of cities or checkerboard squares), then any continuous distribution could at best be an approximation.  The model is based on a process (of choosing squares, or choosing to stay or leave) and the mathematics follows as descriptive; the process is the driver, not the math.

 And yet we need a functional form to which real data can be fit, to be able to estimate component values like the power law exponent and extrapolated limits of city rank and city size.  Recall that all of the distributions begin as power laws and tail-down over time as the checkerboard (or available land) fills up; thus every empirical function has a power law component. Since the model describes a process in time and is easily reduced to an algorithm, simulation provides distributions guaranteed to be appropriate to the model.  Fitting simulation results to a variety of power-law-with-tail functions like
	 $y = Cx^{-r}e^{-(x/\beta)^n}$ and $y = Cx^{-r}(1 - (x/\beta)^n) (n = 0.5...3)$ gives excellent results, with a preference for the parabolic form $(1 - (x/\beta)^2)$ which fits over most parameter ranges, keeps the variable parameter count low, and has a narrative-satisfying hard stop as x \textrightarrow $\beta$.  The reader may experiment with \href{http://scaledinnovation.com/gg/cityGrowth.html}{data fitting and simulation \xlink}.

\section*{Full Range}
\begin{center}
\textcolor{gray}{Insert Figure 6 here.}
\end{center}
 With so much written about city-size distributions over the years, any new theory must cover the entire range of the observed data to be credible. The common issues are illustrated in Figure 6, where the black dots are the data for Swiss municipalities from Decker 2007$^{15}$ and the colored lines are best-fits to three different functions.  It's clear that a pure power law has a prominent ``tail gap": the real data tail down strongly after city rank 1000, but the power law does not.  It's equally clear that the lognormal has a prominent ``head gap", such that several of the largest cities are about three-fold larger than the fit.  The power law*parabola fit is very good over the whole dataset, with a small discrepancy in the high-rank tail -- but then as noted above, this is simply a practical functional choice to impose a tail on a power law, so that the power law component may be extracted and estimated.

 There are more than a few papers that either ignore their inability to account for the full distributions, or else (perhaps in ignorance) use incomplete datasets that exclude the commonly seen high-rank tail.  Krugman himself bases his discussion on the power law apparent in just the 130 largest US cities$^{3}$, which is not nearly enough: the tail-down in the US data is not strongly apparent until about 1950 with more than 4000 cities in the dataset.

\begin{center}
\textcolor{gray}{Insert Figure 7 here.}
\end{center}

    Figure 7 shows that the high-rank tail-down is not unusual; in fact it is the norm for datasets large enough to include cities below about 10,000 population$^{1}$.  Datasets for smaller countries tend to be more complete, thus Switzerland, Finland, and Venezuela show the tail very completely, and so also have clearly extrapolated maximum ranks (2500, 430, and 220, respectively).  The available data for Germany includes its largest 400 cities, the smallest of which has population 34,000, which is not nearly enough to show a tail -- its power law is excellent over a 100-fold range.  The United States 2022 data begins to tail down, a trend that is much more apparent over time (see below).  A subjective assessment of the available national distributions \href{http://scaledinnovation.com/gg/cityGrowth.html}{\xlink} shows 47\% with a strong high-rank tail-down, 15\% essentially pure power law, 23\% distorted by a single very large city, and 17\% with too-small or ill-formed distributions.  For the 50 US states the assessments are 32\%, 34\%, 24\%, and 10\%, respectively.

    Note also in Figure 7 that the power law slopes (solid lines) are similar with values of 2.27, 2.30. 2.27. 2.19, and 2.40 for the United States, Germany, Venezuela, Finland, and Switzerland, respectively, in all cases with fitting uncertainties below 0.01$^{16}$.

    There are two camps in the functional form debate: power law vs lognormal.  Gibrat's law of proportionate growth is sensible: things like cities or economic entities grow proportionately, i.e. multiplicatively, over time and this leads to lognormal distributions$^{17, 18}$.  The fact that many full datasets exhibit a high-rank tail, which is also true for lognormals (Figure 7, in red), supports that model$^{19}$.  But the lognormal has its problems, namely that it is not scale free (contrast to Figures 7, 9) and when fit to the high-rank end of the distribution, it fails at the low-rank (the head gap of Figure 7). The current model offers an alternative to account for the tail, namely the Bagrow model with a finite space.  

\section*{Dynamics}
\begin{center}
\textcolor{gray}{Insert Figure 8 here.}
\end{center}

    Most the literature models do not deal with dynamics.  They ignore time, assume a steady state, assume a fixed number of cities, or initialize their model in an arbitrary state.  These are major flaws since city growth is clearly dynamic and tied to population growth, and especially curious for those who invoke power laws, since the essence of their generation is positive feedback in which the future depends on an earlier state.  Time-dependence is baked into pareto processes.

    Time is intrinsic in the present model since it is process-driven.  It begins with an empty landscape, and as long as there is empty space remaining, growth is pure pareto.  Figure 8 shows multiple time-slices of a simulation with $\alpha$ = 2.5 (thus $r = p(stay) = 2/3$) and a limit of 100 cities. As time proceeds the distributions tail strongly down (because they are anchored) and the apparent power law slope goes from that of the driving model to larger values, in this case from 2.5 to about 3.1; this is the slow approach to a gaussian asymptote (in a log-log rank-frequency plot, a gaussian distribution in the dependent variable appears as a horizontal line with a slight tail up at low rank and tail down at high rank).

\begin{center}
\textcolor{gray}{Insert Figure 9 here.}
\end{center}

    The United States is an excellent laboratory in which to study city growth because it begins with a clean slate (from the European immigrants' point of view, who rarely acknowledged Native American precedence) and census records are complete after 1790.  Figure 9 shows US city distributions from 1790 to 2022 (solid lines) and the power law*parabola function (dashed lines), all with the same $\alpha$ = 2.10 and $\beta$ = 8000, varying only the scale factor (the size of the largest city).  These data validate the current model: the distributions are scale-free (as only power laws can be) with the onset of saturation appearing only at later times as predicted.  Other datasets also show that the distributions are scale-free over substantial time, for example English cities from 1086 to the present, though they are not complete enough to show saturation.
    
\section*{Initialization}
    As noted above, this model implicitly self-initializes.  No additional assumptions about city numbers or distributions are necessary.  Because the premise of next-generation choice to stay-or-leave is personal and local, the model is independent of technology.  Formal ownership of land or resources reinforces the model (a child may inherit the farm or small business), it is not necessary; early agrarian families would occupy and work the same land for generations despite having no formal ownership.  In fact the model may have begun in the Neolithic with the transition to farming and the appearance of permanent settlements.  Human behavior does not change: stay close to family, stay with what you know, don't take unnecessary chances.
    
\section*{Limits and Exceptions}
    The central premise of this proposal is that each generation's stay-or-leave decision drives the model.  The decision drivers of risk- and loss-avoidance should decrease with improved travel and communication: ``My hometown is all I know and it might be worse elsewhere" becomes ``I hear that it's livable over there."  Figure 3 and Kaplanis figures S18 and S19$^{7}$ show when this could have occurred: in North America the birth-death distance increased ten-fold in two generations from those born in 1700 to 1760, and in Europe the change occurred from about 1850 to 1880.   Certainly most cities and towns were nucleated by that point.

    This suggests that while the model could have less influence in the modern era, it may have been dominant for thousands of years before.  Some literature models initialize with a fixed set of cities having normally-distributed sizes; a better initial state would be a power-law distribution left in place by centuries of personal, local decisions.  The model may still apply but urban growth may now additionally depend on modern factors; a modern-era combination of personal-stay-or-leave and economically-driven proportionate growth$^{17}$ is perfectly reasonable -- and difficult to dissect since both give rise to similar distributions.
\begin{center}
\textcolor{gray}{Insert Figure 10 here.}
\end{center}
 
    A subjective assessment of the quality of fit of this model's truncated power law to 158 world and US state datasets is given in Figure 10: 63\% give good to excellent fits with or without a high-rank tail.  About 15\% have small or unusual datasets that inconsistent with any model.  The remaining 22\% all have a small number of cities -- usually one, sometimes two or three -- with sizes a factor of two or more larger than the rest of the distribution would predict.  These are primate cities$^{20}$ like Vienna, Bangkok, and London.  Primate cities are larger and more frequent internationally than in US states; in the US, New York City and Boston have the largest anomalies but are only two-fold larger than expected in their states, while the largest cities in Austria, Thailand, Chad, Uganda, Costa Rica, Uruguay, Vietnam, Georgia, Chile, Mali, Indonesia, and Latvia are all three to fifteen fold larger, as can be \href{http://scaledinnovation.com/gg/cityGrowth.html}{explored here \xlink}.  Primate cities are artificial (in the sense of Herbert Simon) and owe their unusual status to political and economic factors; they fall outside of the current discussion.  The accompanying software can optionally redistribute the population of the largest city, thereby removing the artifact, after which most of these cases are nicely fit by tailed power laws with $\alpha = 2.0 - 2.4$.

\section*{Clear about Alternatives}

    The simplicity of this model and its ability to account for the full range of most city-size distributions challenges the validity of other models.  In particular, models which go to great mathematical lengths to finesse the difference between alternatives (i.e. power law vs lognormal$^{21}$), and yet which fail to account for observed head- or tail-gaps (Figure 7)$^{15}$ can be discarded.   Also models which lack any initialization mechanism, or rely on a hand-waving initial state, can be discarded, or at least relegated to apply only after a different process has set the stage for them.
    
    The clear survivor is the Simon model$^{3, 22}$.  As Krugman notes, ``Simon’s model represents a big improvement over the `economistic' models" because unlike those it predicts a power law.  Whether a power law applies is not a matter of debating fine points of mathematics or data-fitting; the data of Figures 7 and 9, and the fact that the power law is the only scale-free distribution, clinch the case.  The Bagrow model used here is an extension of the Simon model to a finite realm.  What we add is the narrative: the exponent \textit{is} about the probability of forming a new city (per Krugman), but in the sense that new-settlement formation follows from many individuals' stay-or-leave decision.  The value of the mysterious exponent follows directly from the stay-or-leave probability.

    A narrative can also exclude a model.  Those which depend on post-industrial economics or technologies have the problem that power law city size distributions appeared thousands of years ago (vide infra).  These papers' issues may be relevant for today's cities, but a modern mechanism cannot account for ancient demography. 

\section*{Additional Observations}
    To recap: the strength and validity of this model is established by (i) its simple driving process of each generation's stay-or-leave decision, (ii) the quantitative support for that process in millions of genealogical records, (iii) the direct application of that process to the Bagrow finite-space pareto model and its mathematics, which predicts power laws at early times which develop high-rank tails as the space saturates, (iv) the observation of that behavior over the entire range of a majority of accessible city-size datasets, as well as the behavior over time of US cities.

    Given that, it is appropriate to ask what questions this model raises and what it predicts that may be new or unexpected.
    
\section*{Mapping to History}
    For most of history in non-nomadic cultures, population growth has been slow (TFR slightly above 2) and children stayed close to home.  But exceptional circumstances -- war, crop failure, persecution and exile, large-scale colonial opportunism -- may catalyze migration.  The colonial origins of large port cities like Boston, Sydney, and Buenos Aires are well known.  Can genealogical datasets identify examples of towns seeded not by local organic growth (as in this model) but by very specific events, such as the 17th century Ulster Plantations, Palatine migrations of 1708-1710, Irish potato famine, and others?  Are there signatures that distinguish such event-driven seeding of towns from more gradual local expansion seeding?

\section*{Primate Cities}
    The case has been made that primate cities owe their exceptional size to specific circumstance: founding by a natural harbor or crossroads, early political or military influence$^{20}$.  But are there some cases in which a primate city is really just an urban area that has spread to engulf other nearby cities and wrap them all in a single name?  In other words, are there cases in which a historic examination of maps and data could legitimately split a primate city into constituent sub-cities which better fit this generative model?  Are there counterexamples, in which an exceptionally large city grew in geographic isolation from other centers?
    
\section*{Dependence on Net Fertility}
    High net fertility may lead to exhaustion of local resources (i.e. no more available land) and a more urgent need to leave the family homestead. From 1650 to 1880 the United States displayed an exceptional TFR of about 4.8$^{23}$. In a very simplistic model, the first two children inherit the homestead (allowing for free intermarriage, thus balancing out among nearby families) while all other adult children must leave; then $p(stay) = r = 2/$TFR, and since $\alpha$ = 1 + 1/r, $\alpha= 1 + $TFR/2.   Is there evidence to support this relationship?

\section*{Mapping to Specific Families}
    Figure 2 is a  cartoon of a stick figure leaving the homestead.  Given access to massive online genealogies, can we find specific histories -- by the thousands -- that corroborate or refute this picture?  Limited searches are encouraging  \href{http://scaledinnovation.com/gg/gg.html?rr=frontierDemographics}{\xlink}  -- Gerard Spencer's descendants concentrated in Haddam and Colchester CT, William Brewster's in Plymouth MA and New London CT, and William Simonds' in Shrewsbury and Marlborough MA.  Notably they went to different specific places and not to established cities like Boston or New York, consistent with the assertion that it's more about family than economics.  This is not limited to internal or colonial migration; there are anecdotal cases of multiple families from, for example, one Calabrian town mapping to one American town.  Can this be quantified at scale?  If so, it supports the idea of city growth by family connection rather than abstract economic optimization.  Also in such cases is there a period of time, perhaps two or three generations, in which the descendants of a ``leaver" will stay at the new location, and then start to move again themselves?

\section*{Growth and Inter-City Diffusion}
    The primary goal of this work has been to find a narrative and mathematical model that could explain the classic distributions of city sizes. Assuming, for simplicity, exponential internal population growth and no external migration, the model can be extended to country-scale growth and inter-city diffusion.  In a country with $M$ cities and overall population $P$, in each generation a city of population $N$ will have $N(\lambda - 1)$ local births (where $\lambda$ = TFR/2) of which $N(\lambda - 1)(1 - r)$ are lost by out-migration to other cities (where $r = p(stay) = 1/(\alpha - 1)$).  In each generation the city also gains $(P - N)(\lambda - 1)(1 - r)/M$ people by in-migration from all of the other cities.  This is easily computed in a spreadsheet and the results for $\lambda = 1.054, \alpha = 2.4, M = 100, P = 100,000$ are shown below, where city A has initial population N = 10,000, city B has N = 1000 (= P/M, the national average), and small town C has population 10.
\begin{center}
\textcolor{gray}{Insert Figure 11 here.}
\end{center}


    In Figure 11-left, city B, starting at the national average, grows at the same rate as the overall population, as expected.  City A grows but more slowly since it loses more emigrants than it gains immigrants, and vice-versa for small town C.  Over very long time all cities converge to the same size, consistent with Bagrow \textit{et al}'s convergence to a gaussian distribution.

    Figures 11 center shows that for large city A, immigration is minor compared to internal growth and emigration, while in Figure 11 right, small town C is dominated by immigration which exceeds internal growth for 20 generations.  Note that $\lambda$, $\alpha$, and M are realistic values for much of the world, and 100 generations is 3000 years.  Interesting as this is, it is unlikely that real data can be brought to bear:  the United States has perhaps the best level of detail and timespan for census data, but that spans only 1790 to the present, a mere 8 generations.

\section*{Value of the Exponent}
\begin{center}
\textcolor{gray}{Insert Figure 12 here.}
\end{center}

    Krugman$^{3}$ breaks the discussion into three helpful parts: (1) Why are city sizes so diverse? (2) Why do their sizes obey a power law?, and (3) Why is the exponent r $\approx$ 1 (which is to say $\alpha \approx 2$)?.  He notes that $\alpha = 2.0$ implies that everyone must live in one enormous city; mathematically this is a critical point, which seems awkward for a robust natural process.  Here, of course, the exponent is not magical, it is just the probability of the next generation staying or leaving.  As such $0 \leq r \leq 1$, thus $2 \leq \alpha \leq \infty$.  Figure 12 shows the distributions of fitted $\alpha$ for world countries $(n = 108, \alpha = 2.20 \pm 0.37)$ and US states $(n = 50, \alpha = 2.18 \pm 0.36)$.  The range of values for US cities over time (i.e. Figure 9) is narrower $(n = 28, \alpha = 2.14 \pm 0.13)$, as expected from a model with fixed parameters evolving over time.  These values correspond to average probabilities $r$ of staying of 83\%, 85\%, and 88\%, respectively.

\section*{Tail or No Tail}
\begin{center}
\textcolor{gray}{Insert Figure 13 here.}
\end{center}

    Some of the datasets show distinct high-rank tails while others are clean power laws over their entire length.  The data are from common sources and there is nothing obvious to differentiate them.  Curve fitting to all gives a wealth of parameters but there are few correlations between them, even for the relatively homogeneous US state datasets.  Figure 13 is the exception with a clear relationship between the extrapolated maximum number of cities $\beta$ and the state's total population; the light blue line is the implied average city size of 30,000 people (which is 10-100 fold higher than the smallest cities in the datasets).  There is some logic to this -- larger states 'need' more cities, and somehow 30,000 is a common average city size -- and yet there are no such correlations with population density or percent urbanization.

    The states with the cleanest tailless power laws are North Carolina, West Virginia, Louisiana, New Hampshire, Delaware, Maine, Oklahoma, Nebraska, and South Dakota.  Those with the clearest and largest high-rank tails are Texas, Connecticut, Utah, Virginia, Florida, New Jersey, and Rhode Island.  It appears that rural, low-population states are more likely to be unsaturated, while larger or older (colonial) states have reached their city-count limit.
    
\section*{A Finite Number of Cities}
    At first the datasets of Figure 7 and implications of the finite-space Bagrow model seem startling:  there can only ever be a particular number of cities and in many cases we're already at that limit.  Is this a dire ``limits to growth" warning? But since ``city" is used here to mean any fixed settlement, from a village to a megapolis, all this shows is that we've already founded and named all possible settlements.   Our bookkeeping is so good that we can count them all and their populations.  Perhaps aside from Antarctica there is no ``free land" anymore, no place where a person can move that isn't already on the census books.  It's startling to see this as a sharp asymptote in a graph, but the reality is not surprising.
    
\section*{Ancient and Medieval Data}
    Most of the economic optimization literature discusses a world with railroads, canals, highways, and communication systems peopled with rational, mobile economic optimizers, whose choices give rise to the distributions that we see.  But archaeologists see a city as a series of layers, with the buildings and pavement simply the uppermost cap on millennia of history.  You can't dig under a European city without finding Roman ruins, and under those, Iron Age, Bronze Age, and even Neolithic settlements.  A map of Roman Britain is quite familiar, only with different city names -- Londinium (London), Eboracum (York), Lindum (Lincoln). The distribution of city sizes in the Roman Empire is a clean power law $(n = 52, \alpha = 2.09 \pm 0.01$), as is that of European cities in 1500 CE $(n = 2020, \alpha = 2.7 \pm 0.01$) and the Bronze Age Middle East$^{24}$. Any theory of city sizes must apply to the ancient and medieval world as well as to the post-industrial, and so also to slow growth rates: from the Neolithic to the Roman era, the global population grew at a rate of 2.2\% per generation or 0.073\% per year, and from the Roman to the Industrial era, about 5.4\% per generation or 0.18\% per year$^{25}$.

\section*{Speed of Approach to a Distribution}
    Most of the literature concerns steady-state models and ignores initialization, which makes it difficult to say much about the fact that the distribution of cities evolved over time, and yet anything we could say about the time-history of each type of model might be useful in assessing its credibility.  The speed with which a proportional growth model shifts from a uniform to a lognormal distribution is a very strong function of the level of variance assigned to its randomly distributed growth rates; through simulation and comparison to real data those variances might be inferred, then compared to the scatter of known growth rates.  It is similarly difficult to put a timeframe to Simon's model:  how big are the ``lumps" as described by Krugman?  How often do they break away?  

    In contrast the current model produces a complete time course, even starting from a single individual.   We know historic population growth rates, distributions of family sizes, and generation times, which suffice to simulate family tree growth at the individual level -- no parameters need to be guessed or back-fitted.  Add in a literal stay-vs-leave decision for each member of the next generation and a genetic model becomes a city growth model; this is implemented in the haplotree simulation which, like the more abstract checkerboard model, produces tailed power law distributions.  In initial studies these become apparent in 10-20 generations (300-600 years), starting from zero, with Bronze Age survival rates (TFR 2.044) and $\alpha$ = 2.05-2.10.
    
\section*{Extending the Simulations}
    Simulation based on the stay-or-leave model reproduces observed distributions, either pure power law or, with a cap on city number, a power law with a high-rank tail, and this serves its primary purpose.  However simulation may be extended to address additional implications of the model, for example:

\underline{Geographic Dependence} : Though people may leave their homestead of necessity or opportunity, they tend to move the minimum necessary distance; see for example evidence from Anglo-Irish surname mapping over time \href{http://scaledinnovation.com/gg/reports/surnameMigration/surnameMigration-6.jpg}{\xlink}, \href{http://scaledinnovation.com/gg/gg.html?rr=surnameMigration#h7}{\xlink} .  Simulation could model this effect, either with a three-circle model (stay home, stay near, go far) or a continuous distance probability function.  The fraction $(1 - r)$ of ``leavers" would simply be split again by distance.  The simulated results could be checked against historic Anglo-Irish census data (substituting ``county" for ``city", and using rare surnames as surrogates for family connections), and also against genealogical data like those of Kaplanis$^{7}$.

\underline{External Immigration} : Though the current model and simulations deal with a fixed area (a country or a US state) and internal growth, there is no reason that external immigration could not be added, either at historically realistic levels, or just as very small percents used for tracing purposes.  Immigrants could be given the same or modified properties -- for example speaking a minority language might raise the probability of staying in the next generation.

\section*{Last Word}
    Although Krugman prefers Simon's model for its simplicity and power-law outcome, as an economist he finds it painful: Simon's model \textit{``is an extremely nihilistic and simplistic model. It supposes that there are neither advantages nor disadvantages to city size."}$^{3}$  That is precisely the point here -- maybe it's not about economics but about the preference for the next generation to stay in familiar surroundings.  The positive feedback that generates the power law isn't about economic optimization but about keeping close to home and avoiding uncertainty, and the growth comes from more births than deaths.  The preferential attachment is not to a city because it's large, but to a family because they're known and trusted. It doesn't matter whether ``here" is a small village or a crowded city neighborhood: it's home.  By the time that economic arguments come strongly into play, towns and cities will have been nucleated by centuries of generational choice.

\section*{References}

\begin{enumerate}
\item 
    City population data sources: 
    US cities: \url{https://en.wikipedia.org/wiki/1850_United_States_census#City_Rankings}
    
    1000 largest US cities, 2013: from \url{https://gist.github.com/Miserlou/11500b2345d3fe850c92}
      
    2024 urban areas: \url{https://simplemaps.com/data/us-cities}  
    2022 \url{https://www.census.gov/data/tables/time-series/demo/popest/2020s-total-cities-and-towns.html#v2022}
    
    Switzerland: \url{https://en.wikipedia.org/wiki/List_of_cities_in_Switzerland}
     
    Texas: \url{https://www.texas-demographics.com/cities_by_population}
    
    historic US: \url{https://github.com/cestastanford/historical-us-city-populations}
    
    world cities: \url{https://simplemaps.com/data/world-cities}
    
    All datasets and source code for data fitting and simulation
    \href{http://scaledinnovation.com/gg/cityGrowth.html}{are available here \xlink}.

\item Auerbach, F. 1913. Das gesetz der bevölkerungskonzentration [The law of population concentration] Petermanns Geographische Mitteilungen 59, 73-76.  translation by Antonino Ciccone at \url{https://www.vwl.uni-mannheim.de/media/Lehrstuehle/vwl/Ciccone/auerbach_1913_translated_with_introduction_March_2021.pdf}.

\item Krugman, P., 1996. Confronting the mystery of urban hierarchy. Journal of the Japanese and International economies, 10(4), pp.399-418. \url{https://pdfs.semanticscholar.org/e8fd/33532aada66286a09eb9b0a3c1bd3c0498ed.pdf}

\item Simon, H.A., 1955. On a class of skew distribution functions. Biometrika, 42(3/4), pp.425-440. \url{https://snap.stanford.edu/class/cs224w-readings/Simon55Skewdistribution.pdf}

\item 
    Benguigui, L. and Blumenfeld-Lieberthal, E., 2007. A dynamic model for city size distribution beyond Zipf's law. Physica A: Statistical Mechanics and its Applications, 384(2), pp.613-627.   \url{https://www.sciencedirect.com/science/article/pii/S0378437107006061}
    
    Berry, B.J. and Okulicz-Kozaryn, A., 2012. The city size distribution debate: Resolution for US urban regions and megalopolitan areas. Cities, 29, pp.S17-S23.   \url{https://www.sciencedirect.com/science/article/pii/S0264275111001363}
    
    Zanette, D.H. and Manrubia, S.C., 1997. Role of intermittency in urban development: a model of large-scale city formation. Physical Review Letters, 79(3), p.523.  \url{https://link.aps.org/pdf/10.1103/PhysRevLett.79.523}
    
    Verbavatz, V., Barthelemy, M. 2020. The growth equation of cities. Nature 587, 397–401. \url{https://doi.org/10.1038/s41586-020-2900-x}
    
    Xu, Z. and Harriss, R., 2010. A Spatial and Temporal Autocorrelated Growth Model for City Rank—Size Distribution. Urban Studies, 47(2), pp.321-335.\url{https://journals.sagepub.com/doi/pdf/10.1177/0042098009348326}
    
    Batty, M., 2008. The size, scale, and shape of cities. Science, 319(5864), pp.769-771. \url{https://www.jstor.org/stable/pdf/20053314.pdf}   \url{https://pdodds.w3.uvm.edu/teaching/courses/2009-08UVM-300/docs/others/2008/batty2008a.pdf}
        Notes that ``our understanding of how cities evolve is still woefully inadequate."  
    
    Sprung-Keyser, B., Hendren, N. and Porter, S., 2022. The radius of economic opportunity: Evidence from migration and local labor markets. US Census Bureau, Center for Economic Studies. \url{https://www.census.gov/library/stories/2022/07/theres-no-place-like-home.html}
        and \url{https://www2.census.gov/ces/wp/2022/CES-WP-22-27.pdf}

\item Kahneman, D., 2011. Thinking, fast and slow. Macmillan.

\item Kaplanis, J., Gordon, A., Shor, T., Weissbrod, O., Geiger, D., Wahl, M., Gershovits, M., Markus, B., Sheikh, M., Gymrek, M. and Bhatia, G., 2018. Quantitative analysis of population-scale family trees with millions of relatives. Science, 360(6385), pp.171-175. \url{https://www.ncbi.nlm.nih.gov/pmc/articles/PMC6593158/} dataset at \url{https://osf.io/gqxhc}

    Long, L., Tucker, C.J. and Urton, W.L., 1988. Migration distances: An international comparison. Demography, 25, pp.633-640. \url{https://www.jstor.org/stable/2061327?seq=8}

\item Charpentier, A. and Gallic, E., 2018. Internal migrations in France in the nineteenth century. arXiv preprint arXiv:1807.08991. \url{https://arxiv.org/pdf/1807.08991.pdf}
        
\item Ravenstein, E.G., 1889. The laws of migration. Journal of the royal statistical society, 52(2), pp.241-305. \url{https://doi.org/10.2307/2979181}

\item In this paper TFR is defined as the number of children to reach adulthood per couple, thus 2.0 implies a constant population.

\item Newman, M.E., 2005. Power laws, Pareto distributions and Zipf's law. Contemporary physics, 46(5), pp.323-351.  \url{https://arxiv.org/abs/cond-mat/0412004v3}

    Mitzenmacher, M., 2004. A brief history of generative models for power law and lognormal distributions. Internet mathematics, 1(2), pp.226-251. \url{https://www.internetmathematicsjournal.com/article/1385.pdf}
    
   A. Clauset, C.R. Shalizi, and M.E.J. Newman, ``Power-law distributions in empirical data" SIAM Review 51(4), 661-703 (2009). \url{https://arxiv.org/abs/0706.1062}

\item Bratti, M., Fiore, S. and Mendola, M., 2020. The impact of family size and sibling structure on the great Mexico–USA migration. Journal of Population Economics, 33(2), pp.483-529. Figure 2 \url{https://air.unimi.it/bitstream/2434/708193/3/GLO-DP-0392.pdf}

\item Jennings, V., Lloyd-Smith, B. and Ironmonger, D., 1999. Household size and the Poisson distribution. Journal of the Australian Population Association, 16(1), pp.65-84.  \url{https://www.jstor.org/stable/41110478}  and  Jarosz, B., 2021. Poisson distribution: A model for estimating households by household size. Population Research and Policy Review, 40(2), pp.149-162. \url{https://link.springer.com/article/10.1007/s11113-020-09575-x}

\item Bagrow, J.P., Sun, J. and ben-Avraham, D., 2008. Phase transition in the rich-get-richer mechanism due to finite-size effects. Journal of Physics A: Mathematical and Theoretical, 41(18), p.185001.  \url{https://arxiv.org/pdf/0712.2220.pdf}  , \url{https://www.bagrow.com/pdf/a8_18_185001.pdf}

\item Decker, E.H., Kerkhoff, A.J. and Moses, M.E., 2007. Global patterns of city size distributions and their fundamental drivers. PLoS One, 2(9), p.e934.
    \url{https://journals.plos.org/plosone/article/file?id=10.1371/journal.pone.0000934&type=printable}  

\item All data fitting is done by Nelder-Mean least-squares optimization, unless otherwise noted to the function $y = Cx^{-r}(1 - (x/\beta)^n)$ where x = city rank, y = city population; fitted are C, $\beta$, and $r = 1/(\alpha - 1)$.  Standard deviations of fitted parameters are estimated from curvature at the error minimum: Phillips, G.R., Eyring, E.M. 1987. Error Estimation Using the Simplex Method in Nonlinear Least Squares Data Analysis.  All source code available at reference 1.

\item Gabaix, X., 1999. Zipf's law for cities: an explanation. The Quarterly journal of economics, 114(3), pp.739-767, Gabaix, X., 1999. Zipf's Law and the Growth of Cities. American Economic Review, 89(2), pp.129-132. \url{https://www.aeaweb.org/articles?id=10.1257/aer.89.2.129}    
     
\item Gibrat, R. 1931 R., Les in\'egalit\'es e\'economiques. Paris, France: Librairie du Recueil Sirey

\item Eeckhout, J., 2004. Gibrat's law for (all) cities. American Economic Review, 94(5), pp.1429-1451. \url{https://www.jstor.org/stable/pdf/3592829.pdf}

\item Jefferson, M. 1939. The law of the primate city. Geographical Review, 29(2), pp.226-232. \url{https://www.jstor.org/stable/209944,}   Ades, A.F. and Glaeser, E.L., 1995. Trade and circuses: explaining urban giants. The Quarterly Journal of Economics, 110(1), pp.195-227.  \url{https://www.nber.org/system/files/working_papers/w4715/w4715.pdf}  , 
    and Krugman, P., 1996. Urban concentration: the role of increasing returns and transport costs. International Regional Science Review, 19(1-2), pp.5-30.  \url{https://www.rrojasdatabank.info/wbdevecon94/247-283.pdf}  

\item Malevergne, Y., Pisarenko, V. and Sornette, D., 2011. Testing the Pareto against the lognormal distributions with the uniformly most powerful unbiased test applied to the distribution of cities. Physical Review E, 83(3), p.036111.  \url{https://journals.aps.org/pre/abstract/10.1103/PhysRevE.83.036111}

\item Berry, B.J. and Garrison, W.L., 1958. Alternate explanations of urban rank-size relationships. Annals of the Association of American Geographers, 48(1), pp.83-90. \url{https://doi.org/10.1111/j.1467-8306.1958.tb01559.x}  
    ``If we were to use the dictum of the simplest of alternate hypotheses, the selection of explanations would be clear. Instead of the troublesome theories of Zipf, Rashevsky, and Christaller, one would rely for explanation of city-size regularities on the implications of the work of Simon."
    ``The theory still leaves a central question unanswered, namely: Why is the arrangement of city sizes the outcome of simple probabilistic processes?" 

\item Haines, M.R., 1994. The population of the United States, 1790-1920. National Bureau of Economic Research, historical paper 56. \url{https://www.nber.org/system/files/working_papers/h0056/h0056.pdf}    Haines cites 3\% annual population growth; at 30 years per generation, this gives $TFR = 2(1.03^30) = 4.8$, i.e. on average a couple would have nearly five children reach adulthood.

\item Russell, J.C., 1958. Late ancient and medieval population. Transactions of the American Philosophical Society, 48(3), pp.1-152. \url{https://www.jstor.org/stable/1005708} compiled in Wikipedia \url{https://en.wikipedia.org/wiki/Demography_of_the_Roman_Empire}, and Barjamovic, G., Chaney, T., Coşar, K. and Hortaçsu, A., 2019. Trade, merchants, and the lost cities of the bronze age. The Quarterly Journal of Economics, 134(3), pp.1455-1503. \url{https://academic.oup.com/qje/article/134/3/1455/5420484?login=true}, Buringh, E., 2021. The population of European cities from 700 to 2000: Social and economic history. Research Data Journal for the Humanities and Social Sciences, 6(1), pp.1-18. \url{https://brill.com/view/journals/rdj/6/1/article-p1_3.xml?language=en}

\item World population over time: \url{https://www.worldometers.info/world-population/world-population-by-year/}
    
\end{enumerate}

\pagebreak

\begin{figure}
    \centering
    \caption{Auerbach's German cities data}
    \includegraphics*[width=8cm]{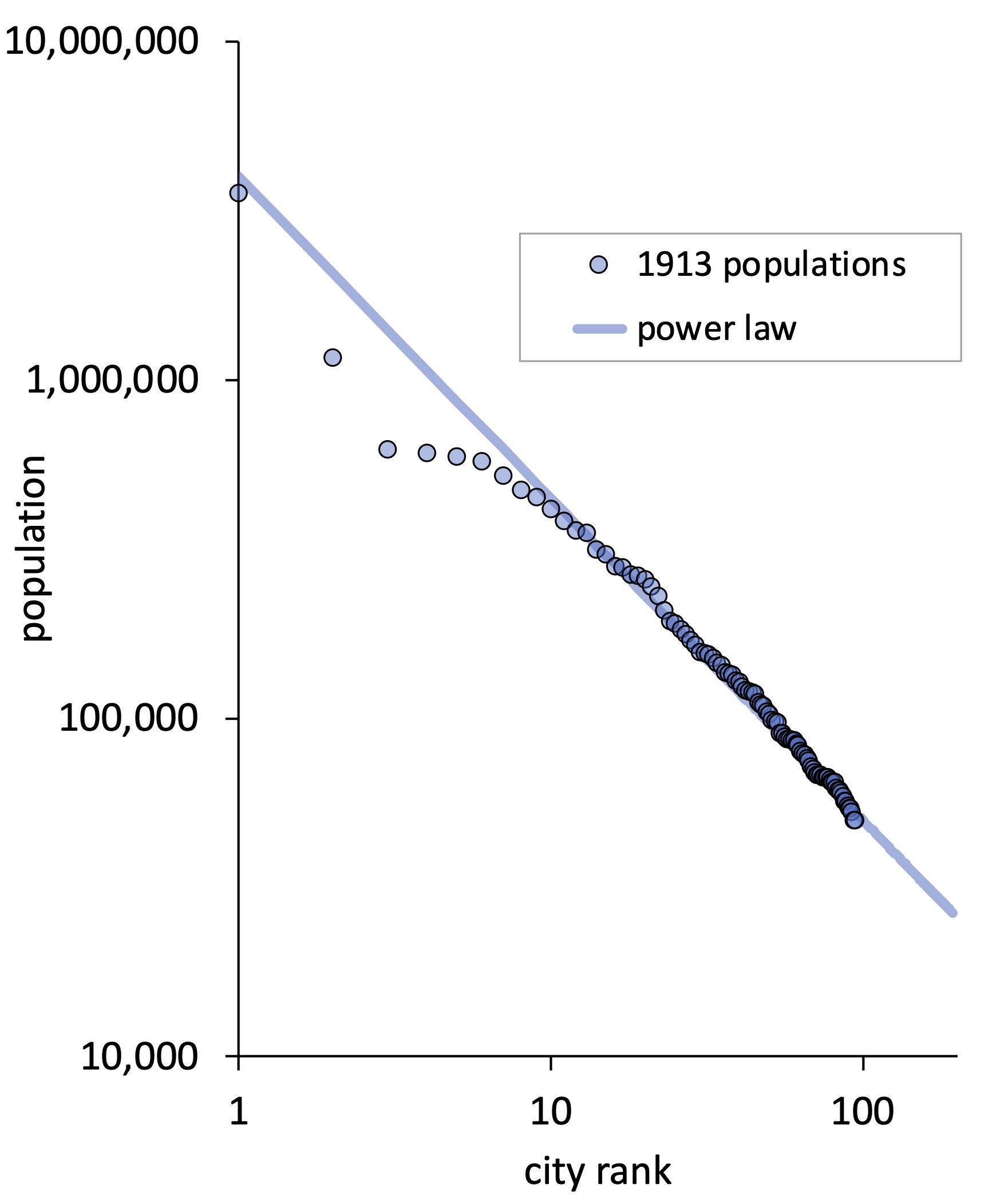}
    
\end{figure}

\begin{figure}
    \centering
    \caption{The stay-or-leave model}
    \includegraphics*[width=8cm]{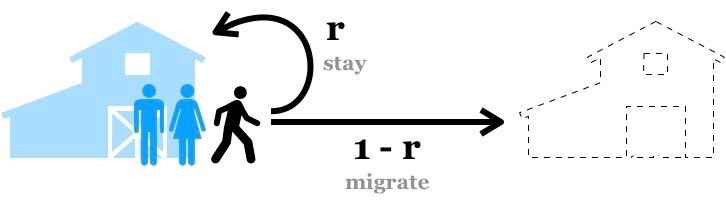}
\end{figure}

\begin{figure}
    \centering
    \caption{Lifetime migration distances from large genealogies}
    \includegraphics*[width=14cm]{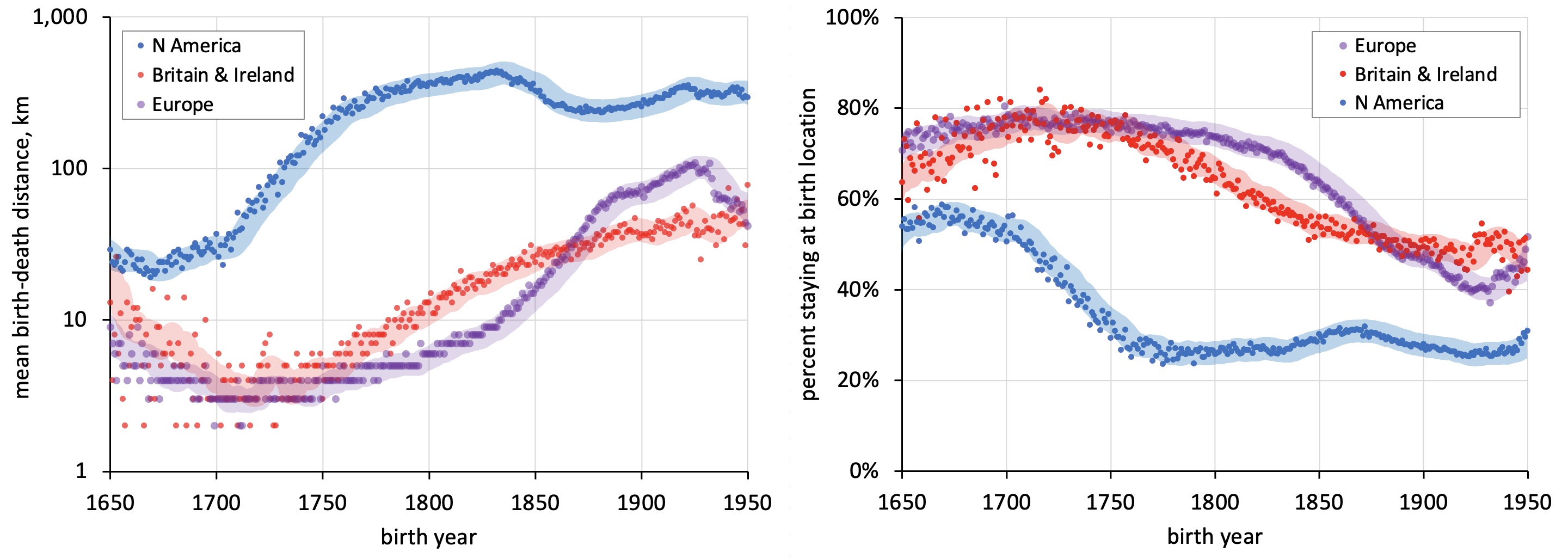}
\end{figure}

\begin{figure}
    \centering
    \caption{stay-or-leave in 19th century France}
    \includegraphics*[width=8cm]{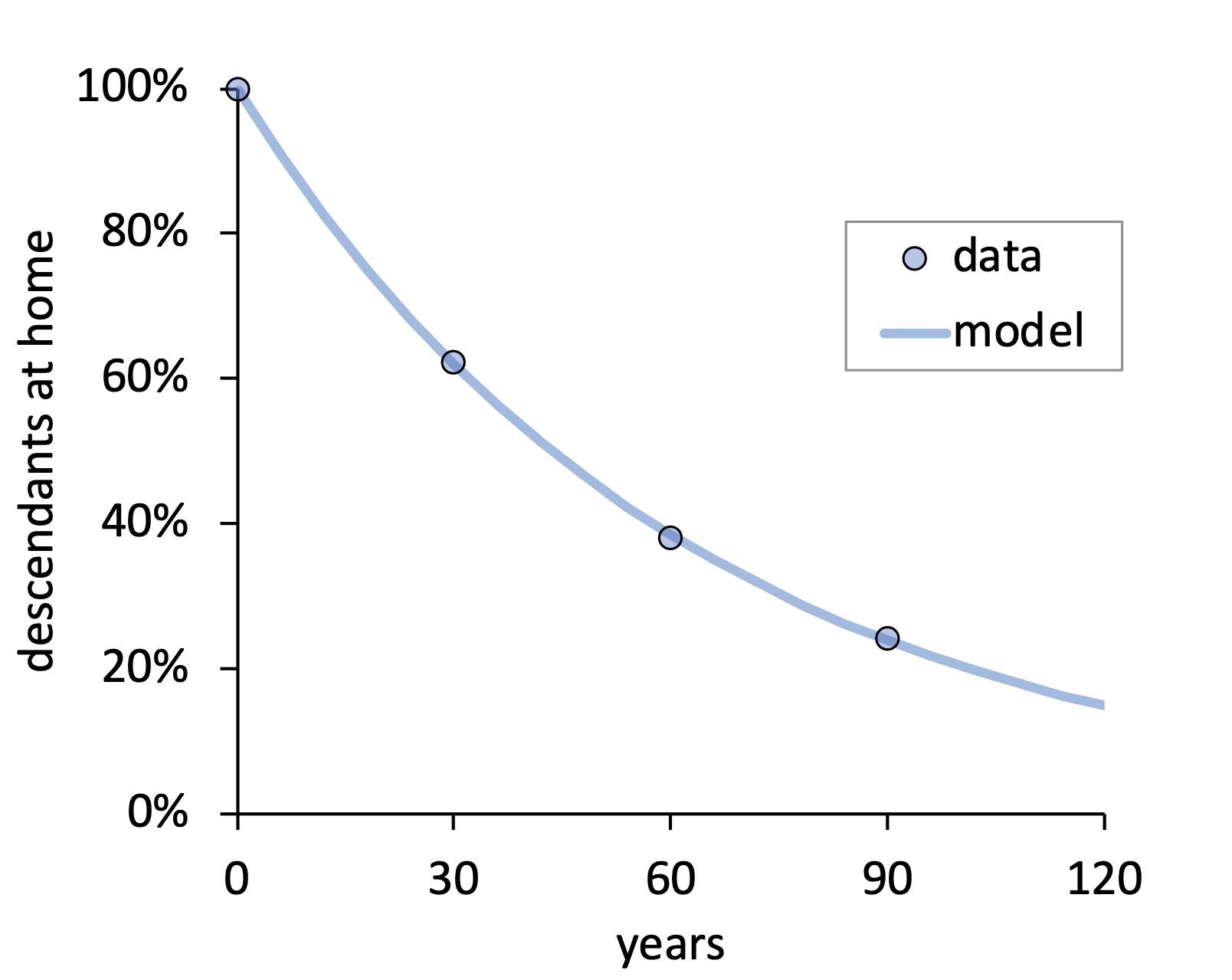}
\end{figure}

\begin{figure}
    \centering
    \caption{The checkerboard metaphor}
    \includegraphics*[width=6cm]{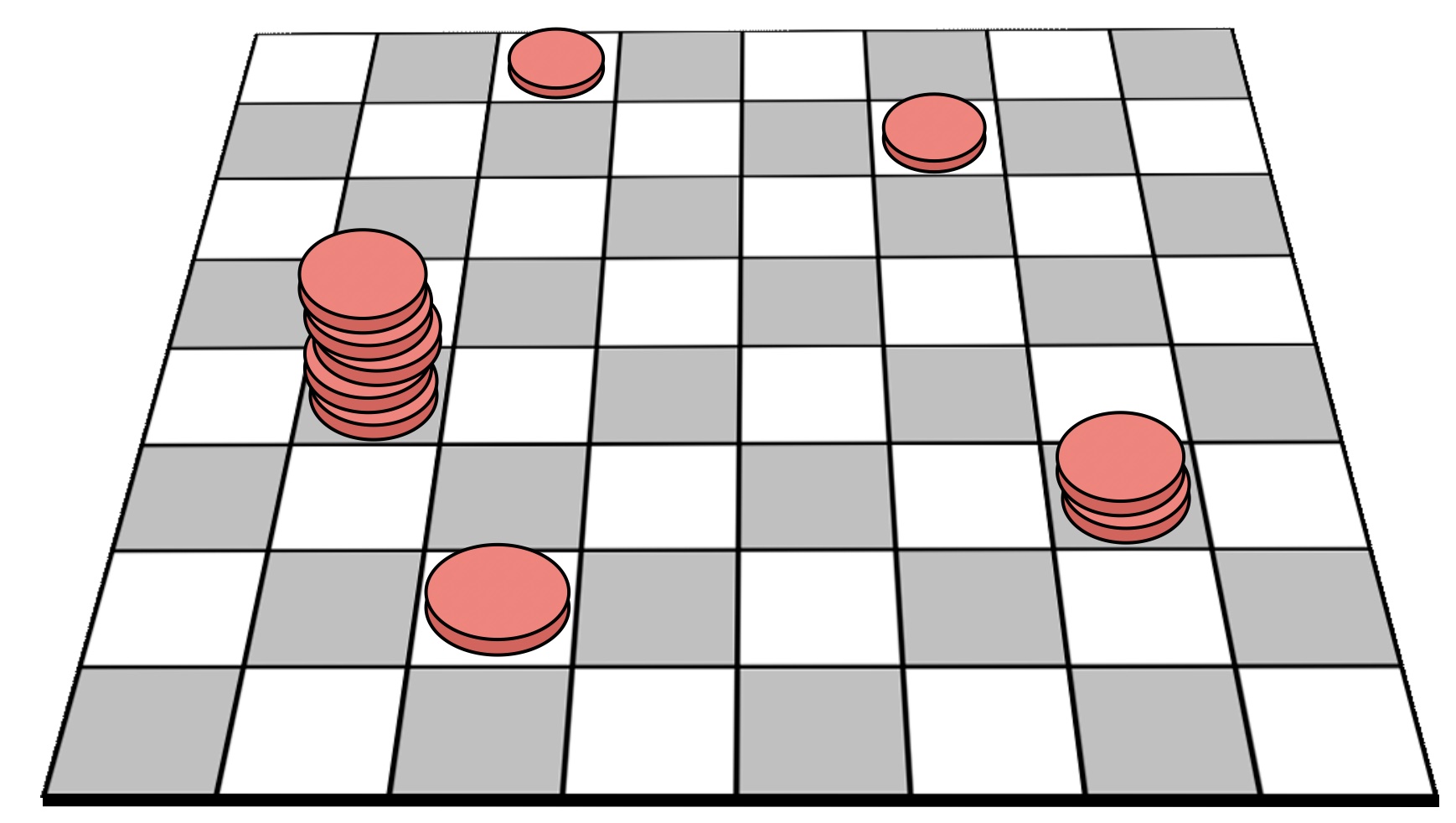}
\end{figure}

\begin{figure}
    \centering
    \caption{Swiss municipalities dataset with common fitting problems}
    \includegraphics*[width=12cm]{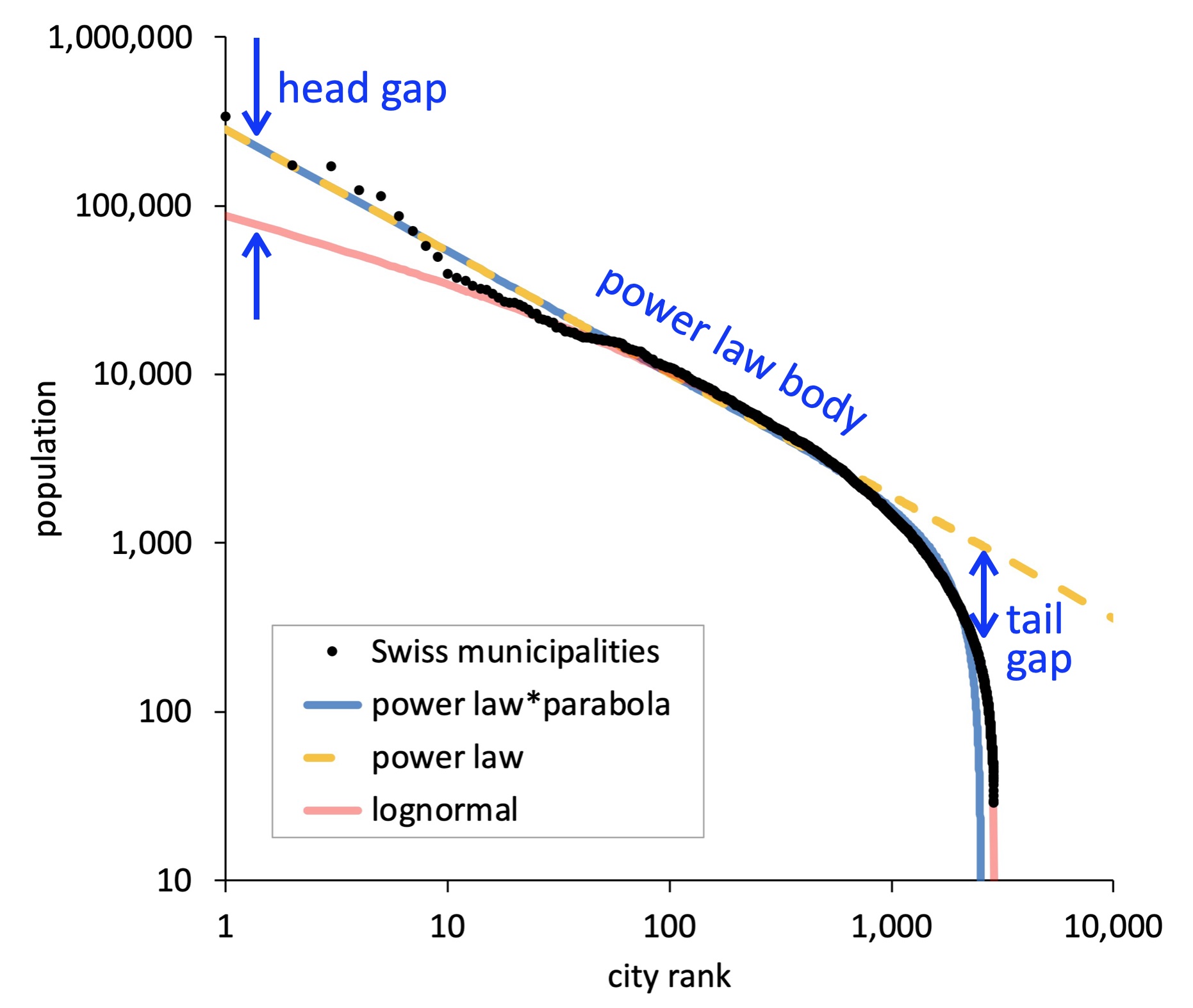}
\end{figure}

\begin{figure}
    \centering
    \caption{National datasets show common slope, some with saturation}
    \includegraphics*[width=12cm]{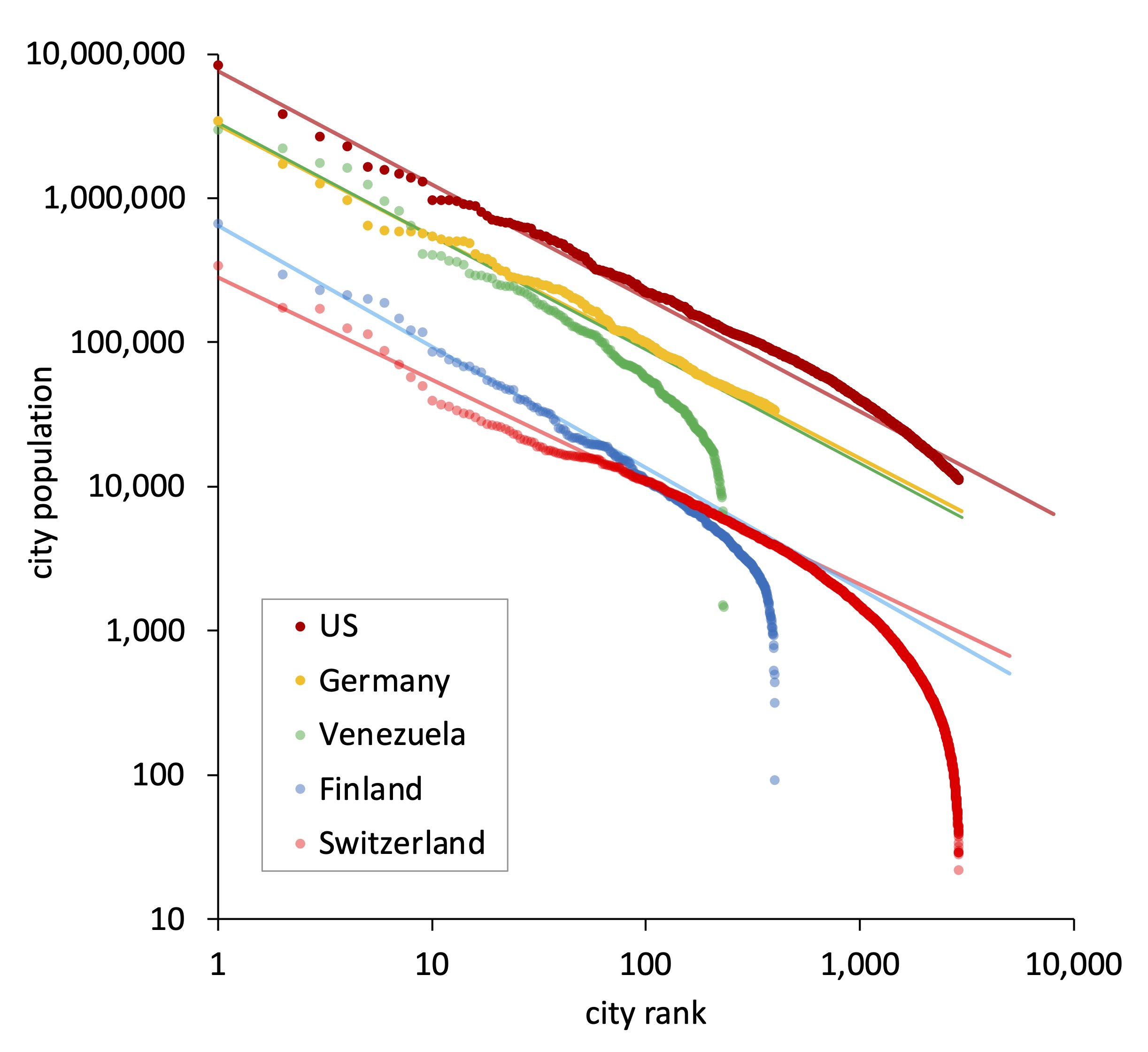}
\end{figure}

\begin{figure}
    \centering
    \caption{Time slices of simulation evolve from power law to saturation}
    \includegraphics*[width=12cm]{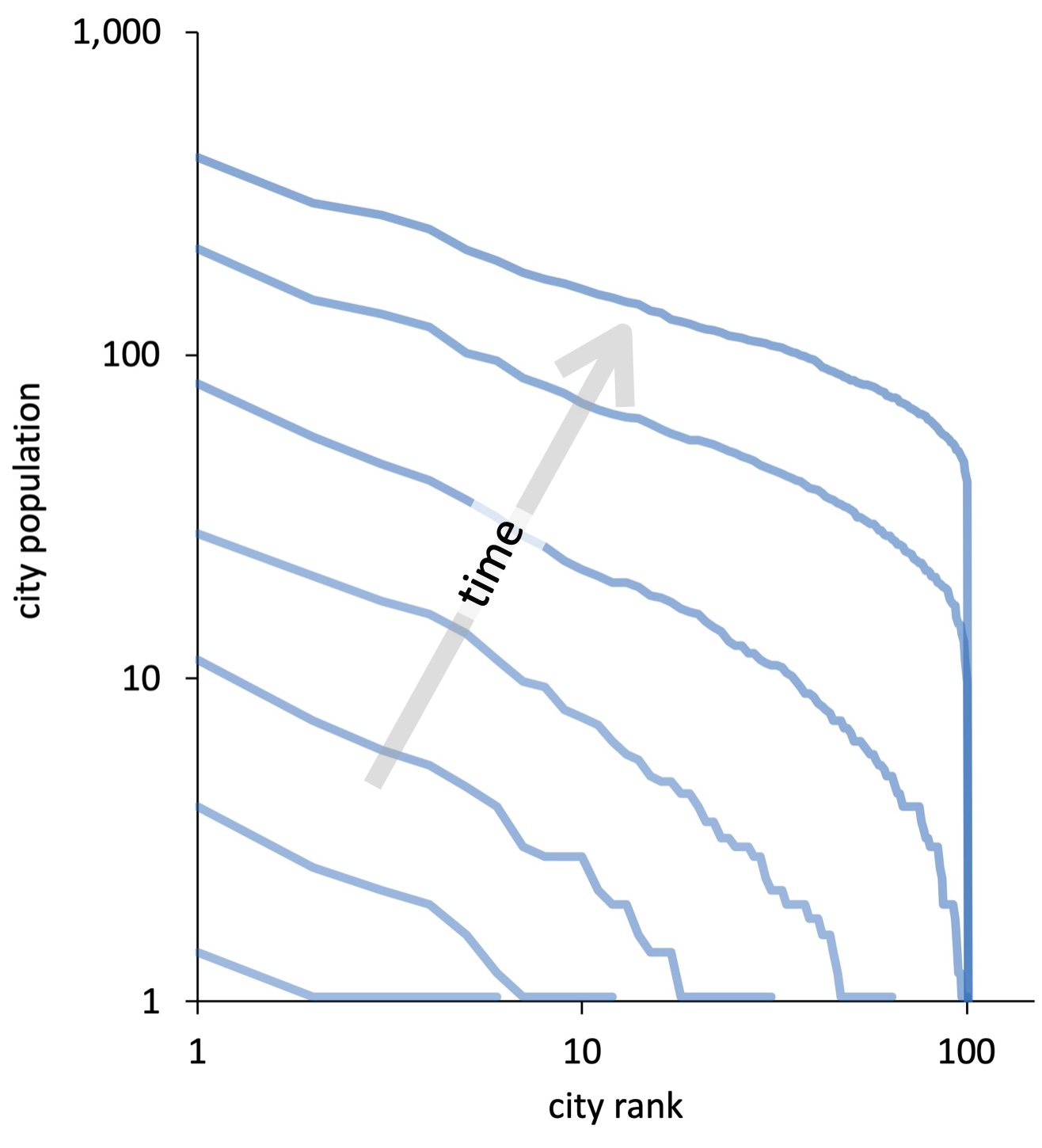}
\end{figure}

\begin{figure}
    \centering
    \caption{Time slices of US city sizes 1790-2022}
    \includegraphics*[width=12cm]{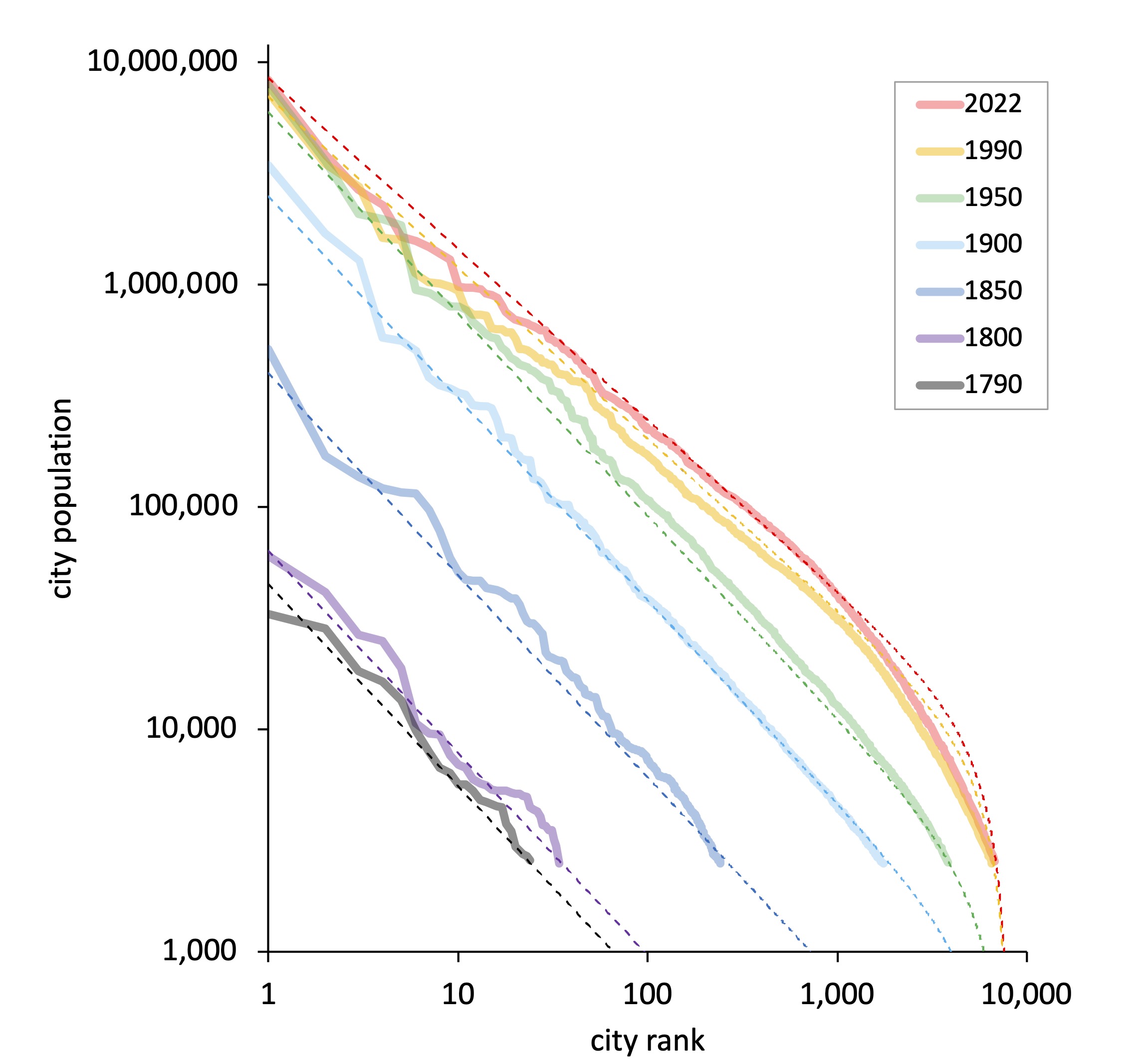}
\end{figure}

\begin{figure}
    \centering
    \caption{Subjective assessment of national and state datasets}
    \includegraphics*[width=6cm]{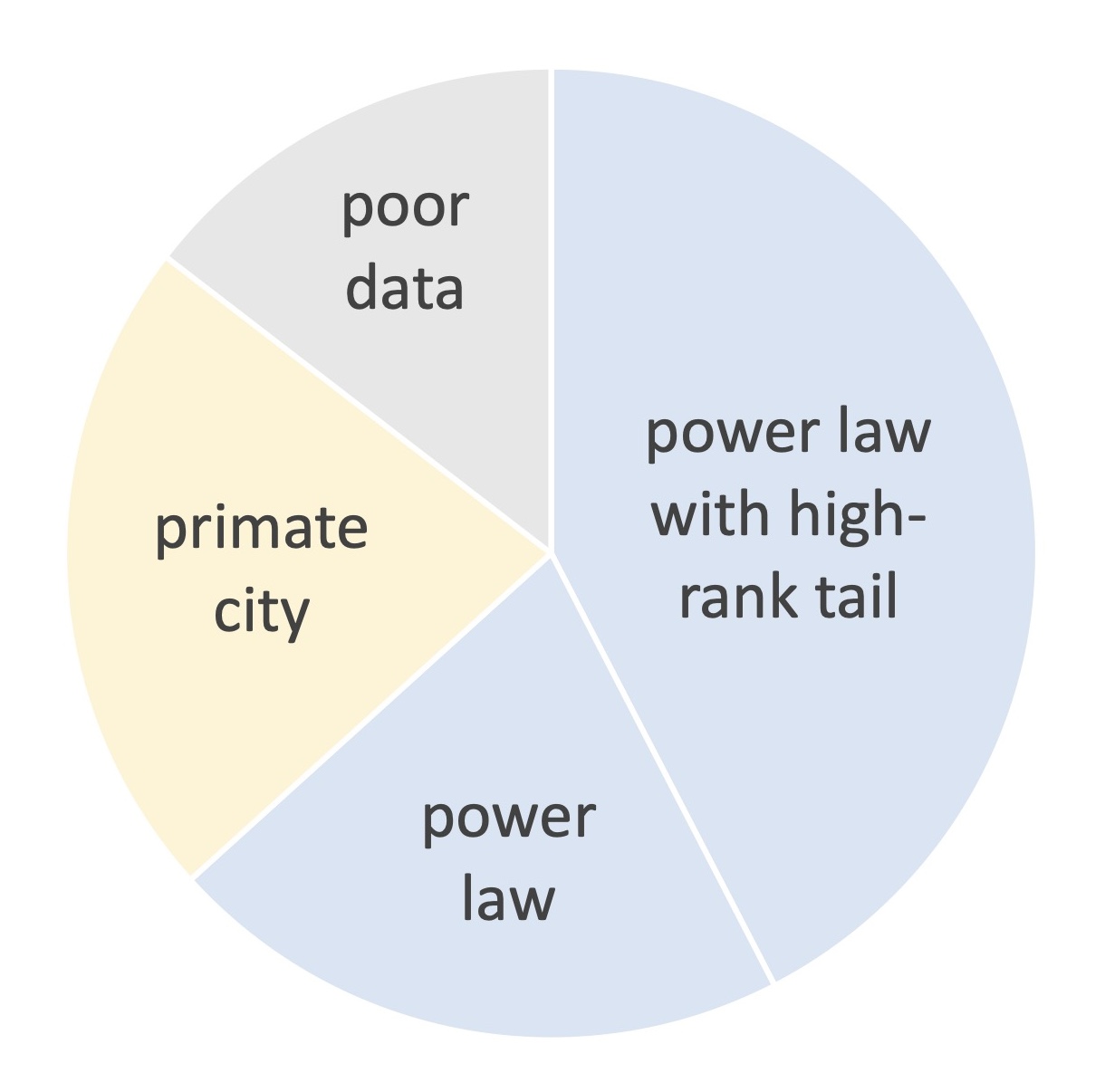}
\end{figure}

\begin{figure}
    \centering
    \caption{Model extension shows inter-city size and migration effects}
    \includegraphics*[width=12cm]{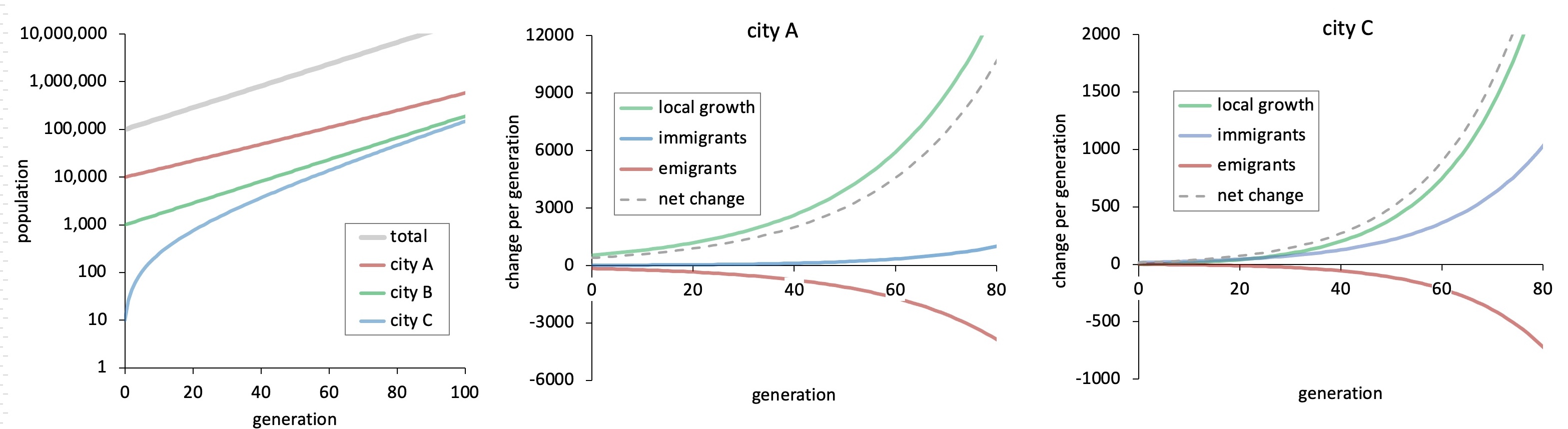}
\end{figure}

\begin{figure}
    \centering
    \caption{Distributions of observed power law exponents}
    \includegraphics*[width=8cm]{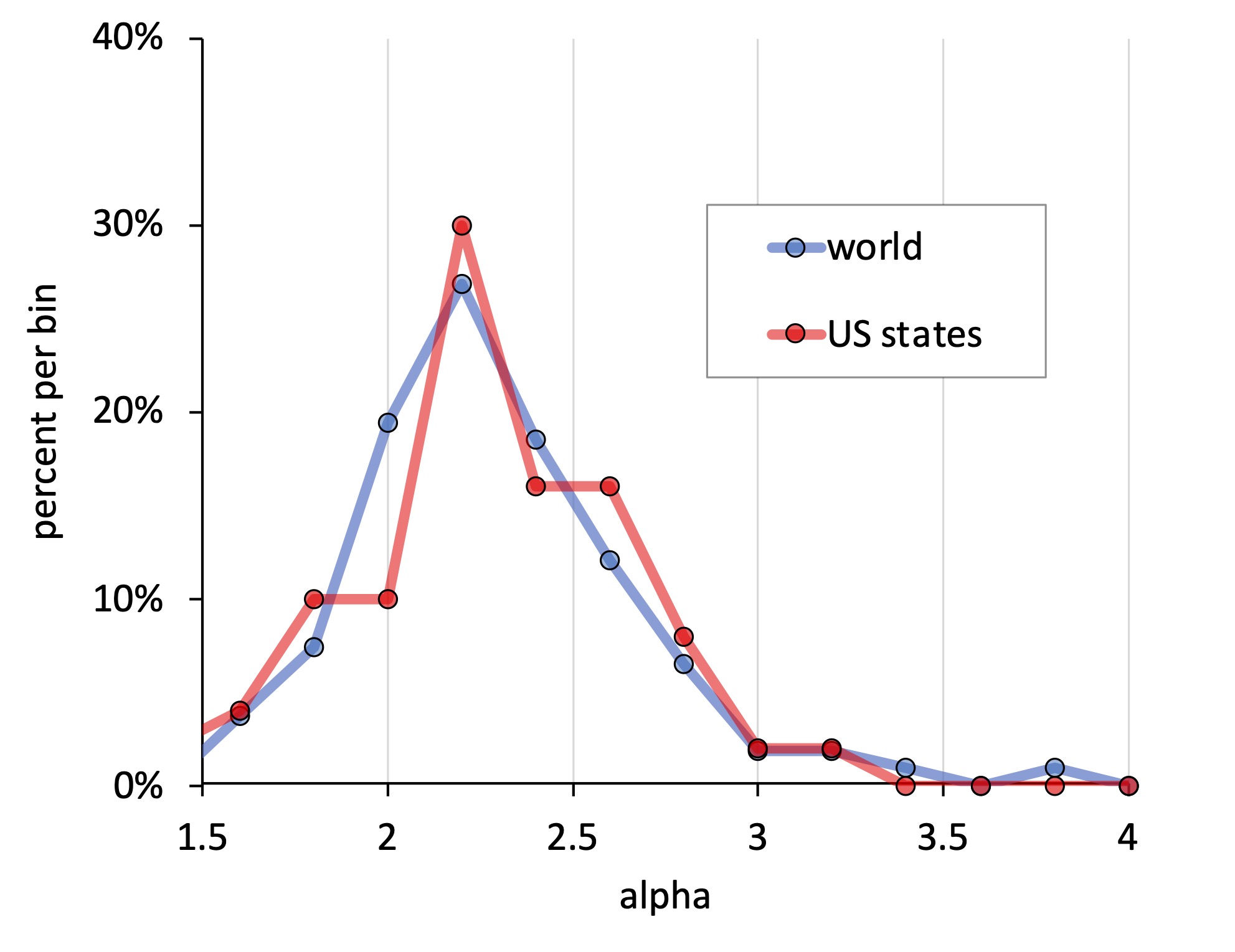}
\end{figure}

\begin{figure}
    \centering
    \caption{Correlation of US state population with maximum city count}
    \includegraphics*[width=8cm]{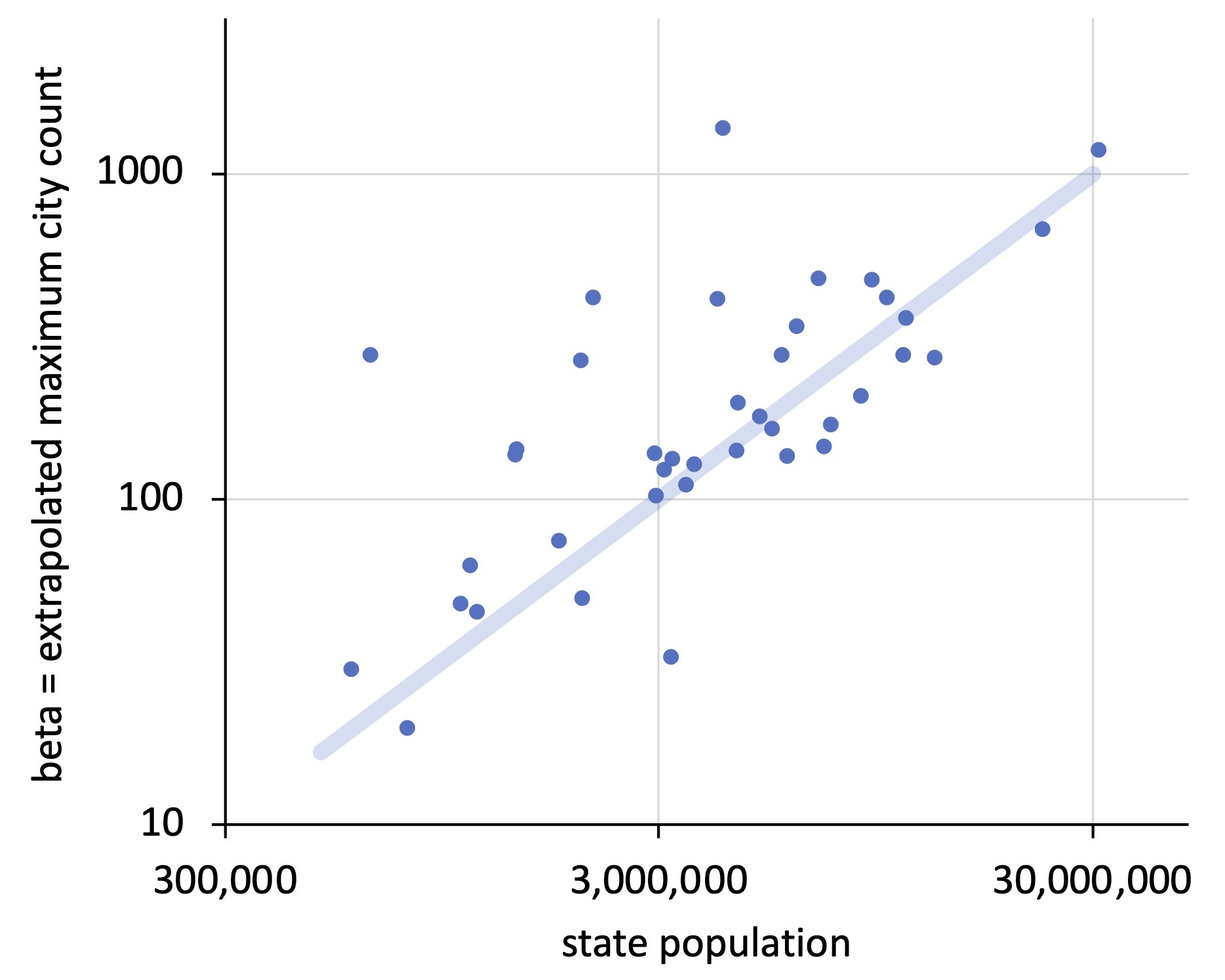}
\end{figure}

\end{document}